# Characterization of Fully Depleted CMOS Active Pixel Sensors on High Resistivity Substrates for Use in a High Radiation Environment

Toko Hirono, Marlon Barbero, Patrick Breugnon, Stéphanie Godiot, Tomasz Hemperek, Fabian Hügging,
Jens Janssen, Hans Krüger, Jian Liu, Patrick Pangaud, Ivan Perić, David-Leon Pohl,
Alexandre Rozanov, Piotr Rymaszewski, and Norbert Wermes

*Abstract*–Depleted CMOS active sensors (DMAPS) are being developed for high-energy particle physics experiments in high radiation environments, such as in the ATLAS High Luminosity Large Hadron Collider (HL-LHC). Since charge collection by drift is mandatory for harsh radiation environment, the application of high bias voltage to high resistive sensor material is needed. In this work, a prototype of a DMAPS was fabricated in a 150nm CMOS process on a substrate with a resistivity of >2 kΩ·cm that was thinned to 100 μm. Full depletion occurs around 20V, which is far below the breakdown voltage of 110 V. A readout chip has been attached for fast triggered readout. Presented prototype also uses a concept of sub-pixel en/decoding three pixels of the prototype chip are readout by one pixel of the readout chip. Since radiation tolerance is one of the largest concerns in DMAPS, the CCPD_LF chip has been irradiated with X-rays and neutrons up to a total ionization dose of 50 Mrad and a fluence of $10^{15} n_{eq}/cm^2$, respectively.

## I. INTRODUCTION

DEPLETED CMOS active pixel sensors in commercial technologies [1] [2] [3] [4] are an attractive detector choice for high-energy particle physics experiments. The detectors of this type are based on the multiple-well structures of CMOS technology in which the readout circuitry can be within the collection well. Thus, complex electronics can be implemented with a high fill factor. This feature allows for radiation hardness and fast readout, which are required in experiments conducted in harsh radiation environments such as in the ATLAS HL-LHC [5].

The outer layer of the pixel detectors in the ATLAS HL-LHC needs to withstand irradiation to a total ionization dose (TID) of 50 Mrad and fluence of $1\times10^{15}$ $n_{eq}/cm^2$ [5]. Because charges created in the diffusion area of the irradiated chip are not collected efficiently by the collection well [6], therefore charge collection by drift is mandatry. Thinning and the backside process are optional features of the CMOS active sensor that can reduce material budged without negative impact on charge collection.

There are two approaches to fast readout. One is to implement all of the readout chain in the CMOS active sensor, and the other is to attach an existing readout chip to the CMOS sensor [7]. The development of the former has just begun [8] [9]. With the latter approach, fast readout can be achieved without adding much electronics to the sensor. Furthermore, multiple pixels of the CMOS active sensor can be readout by one pixel of the readout chip (sub-pixel en/decoding). A high spatial resolution without fine pitch bump bonding is possible.

In this work, a prototype chip [3] [4] was developed and its radiation hardness was tested. It was also attached to a readout chip, which enabled fast readout. The prototype is described in Section 2. The results of its X-ray and neutron irradiation are discussed in Section 3. The prototype is attached to a readout chip to test fast readout, as discussed in Section 4, and the conclusion of the work is given in Section 5.

## II. PROTOTYPE

CCPD_LF is a prototype chip with a size of 5 mm × 5 mm produced in 150 nm LFoundry technology [3] [4]. The wafer provided by the foundry has a resistivity of >2 kΩ·cm. The breakdown voltage was measured at 110V and the maximum depletion depth was measured to be 170 μm in a previous study [3]. In the present study, the chip was thinned to 300 μm and 100 μm, and an aluminum contact was grown on the back side.

The CCPD_LF chip can be attached the readout chip of the ATLAS pixel detector (FE-I4) [10]. The pixel size of the CCPD_LF chip is 33 μm × 125 μm and the size of 3 × 2 pixels equals the size of 2 pixels on the FE-I4, with a pixel size of 50 μm × 250 μm. Three pixels of the prototype chip are readout by 1 pixel of the FE-I4 chip [7]. The schematic is shown in Fig. 1. The readout circuitry of the CCPD_LF chip consists of a test pulse input, charge sensitive amplifier (CSA), comparator with 4-bit trim DAC, and an output stage. The output stage re-shapes the output pulse of the comparator. The output stages of the three pixels generate current with different

Manuscript received December 9, 2016. This project has received funding from the European Union's Horizon 2020 Research and Innovation programme under Grant Agreement no. 654168. This project supported in part by the IEKP Irradiation Center at Karlsruher Institut für Technologie.

T. Hirono, T. Hemperek, F. Hügging, J. Janssen, H. Krüger, D.-L. Pohl, P. Rymaszewski, and N Wermes are with Physikalisches Institut, Universität Bonn, Bonn 53115 Germany (e-mail: hirono@physik.uni-bonn.de).

M. Barbero, P. Breugnon, S. Godiot, J. Liu, P. Pangaud, and A. Rozanov are with Centre de Physique des Particules de Marseille, Aix-Marseille Université, Marseille 13288 France.

I. Perić is with Institut für Prozessdatenverarbeitung und Elektronik, Karlsruher Institut für Technologie, Karlsruhe, Germany.

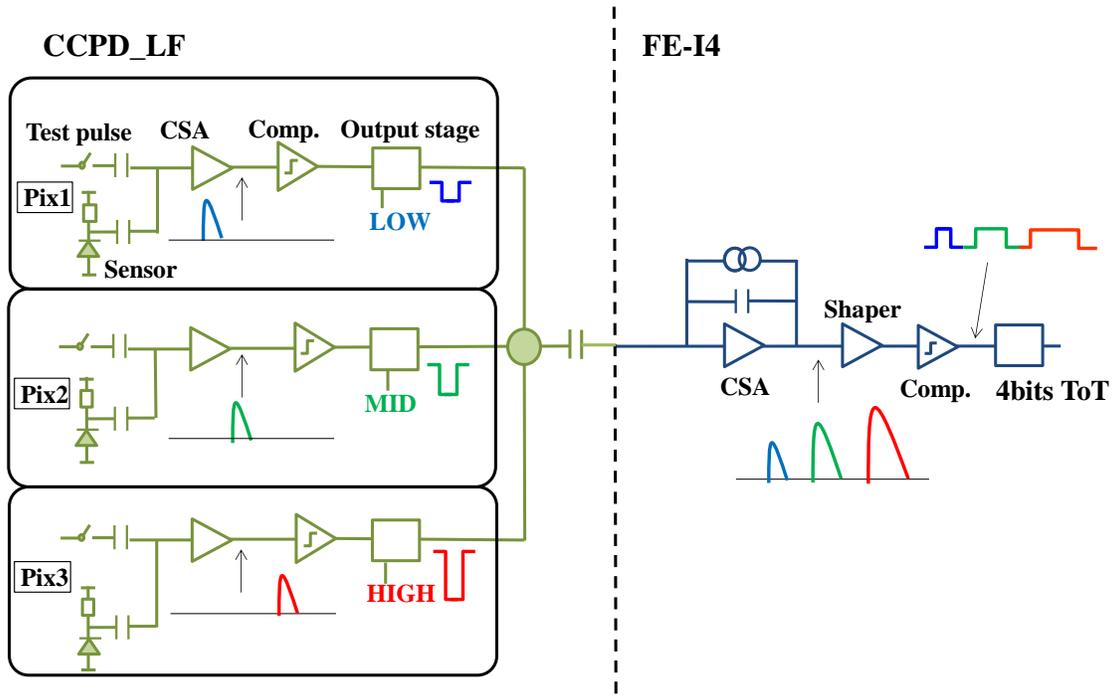

Fig. 1. Block diagram of the readout circuitry of the CCPD_LF chip attached to the FE-I4 readout chip. Three pixels of the CCPD_LF chip are connected to one pixel of the FE-I4 chip. Wave forms of the top pixel (Pix1), middle pixel (Pix2), bottom pixel (Pix3) at each point are shown in blue, green and red, respectively.

amplitudes in the FE-I4 CSA. The amplitude of each output stage is distinguishable with a 4-bit time-over-threshold (ToT) circuit in the FE-I4 chip. In this way, the hit-position is encoded in the CMOS active pixel sensor and decoded in the readout chip. Therefore, the hit position is reconstructed by this sub-pixel en/decoding scheme.

### III. Depletion voltage

A spectrum of the radioactive source, $^{55}$Fe, was acquired to evaluate the full depletion voltage of the prototype, thinned to 100 μm. $^{55}$Fe emits Mn $K_\alpha$ with an energy of 5.9 keV and its attenuation length in silicon is 28.69 μm [11]. When a photon is absorbed in the depletion region of the sensor, most of the charge carriers are collected by a single pixel unless the absorption occurs near the pixel edge. On the other hand, when a photon is absorbed in the diffusion region, more charge carriers diffuse to multiple pixels. Therefore, the peak of the spectrum is higher in a fully depleted sensor than in a partially depleted sensor. X-rays were injected into the back side of the chip to emphasize the difference between the fully depleted and partially depleted conditions of the sensor.

The results are shown in Fig. 2. The slope changes around the bias voltage of 20 V and intensity of the peak saturates above this bias voltage. This shows that the full depletion voltage is around 20 V. This is lower than the break down voltage by a factor of five. The result indicates that the sensitive volume of the CCPD_LF chip can be fully depleted stably.

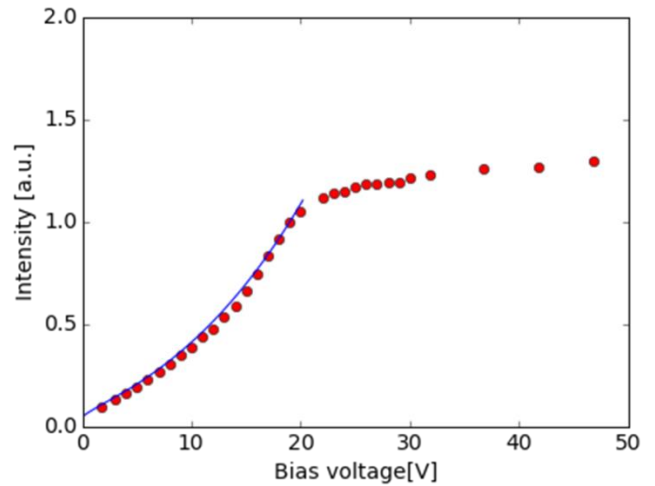

Fig. 2. The peak intensity of the 55Fe spectrum as a function of bias voltage.

### IV. Radiation hardness

#### A. X-ray irradiation

TID radiation hardness was tested using an X-ray tube at the IEKP Irradiation Center at the Karlsruhe Institute for Technology [12]. The CCPD_LF chip has twelve flavors of pexels that can be grouped into three categories with respect to type of transistor used in CSA's feedback (Table 1).
The irradiation was performed at room temperature with a rate of 1.6 Mrad/h stopping at 11 steps for the measurements. A test pulse injection method was used to obtain the gain and noise of the readout. Annealing was performed for 9 h at room temperature at TID of 4 Mrad. Fig. 3 shows the gain and noise of the readout, normalized to the gain and noise before irradiation, respectively. All the readout types are functional after the irradiation.

However, the gains decreased by 20-30% with no significant difference between readout types. With the exception of the CSA feedback transistor, the transistors in the readout circuitry are standard linear type. The measurement result indicates that there are sensitive transistors in the readout circuit other than the feedback transistor of the CSA.

TABLE I. TRANSISTOR IN FEEDBACK OF CSA

| READOUT | TRANSISTOR TYPE | WIDTH/LENGTH |
|---|---|---|
| DEFAULT | LINEAR | 350 µm /0.9 µm |
| LONG | LINEAR | 0.35 µm /1.5 µm |
| ELT | ENCLOSED LAYOUT | 2.43µm/0.16µm |

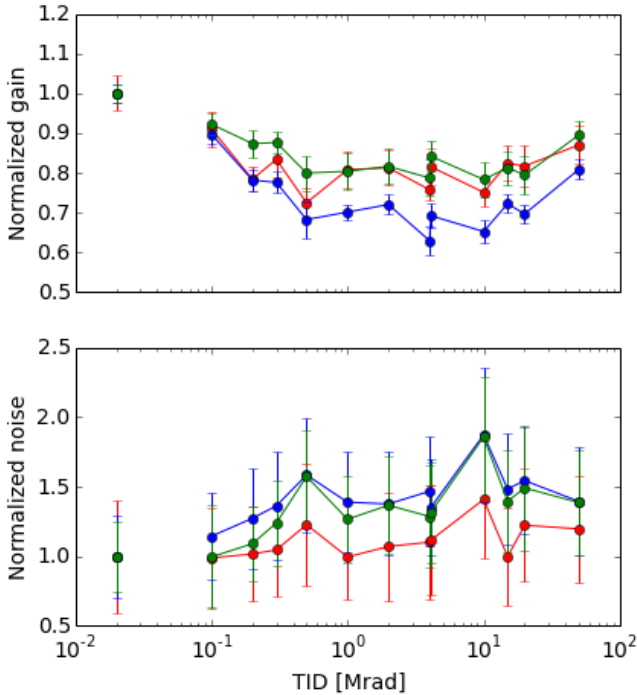

Fig. 3. The normalized gain and noise of CCPD_LF chip. CCPD_LF readout with are default linear transistor (green), long linear transistor (blue), and enclosed layout transistor (red).

### B. Neutron irradiation

Irradiation of the prototype chip with neutrons up to the fluence of $1 \times 10^{15}$ $n_{eq}/cm^2$ was carried out at the TRIGA Mark II reactor at the Jožef Stefan Institute [13].

$^{55}$Fe and $^{241}$Am spectra from the non-irradiated and irradiated chip are shown in Fig. 4. The thickness of the chips is 300 µm. The bias voltage of the non-irradiated and irradiated chip was 100 V and 125 V, respectively. A peak at 5.9 keV and 59.6 keV is clearly seen in the spectra of non-irradiated and irradiated chip. However, the signal amplitude of the irradiated chip is smaller than that of non-irradiated chip. The readout gain of the irradiated chip measured using test pulse injection is smaller than that of the non-irradiated chip by the factor of 0.65. The decrease of the signal amplitude is mainly caused by of the degradation of the readout gain, which can be understood by the background of

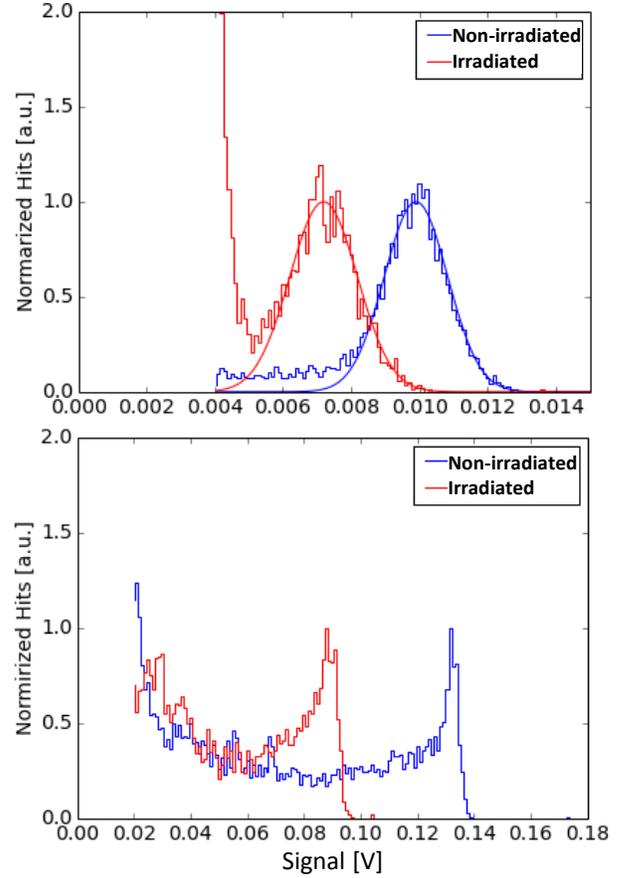

Fig. 4. Spectra of $^{55}$Fe (upper graph) and $^{241}$Am (lower graph) in the non-irradiated and irradiated chip. The spectra are normalized to the peak at 5.9 keV and 59.6 keV, respectively.

the reactor, which had an estimated TID of 1 Mrad. In addition, the fact that the non-irradiated and radiated chips are different devices may explain the difference in their gain.

## V. SUB-PIXEL EN/DECODING

The concept of the sub-pixel en/decoding was tested. CCPD_LF was bump-bonded to the FE-I4 readout chip and the FE-I4 chip was tuned with the standard tuning procedure of the readout software, Bonn ATLAS Readout in Python and C++ (PyBAR) [14]. After the standard tuning, the feedback current of CSA and the threshold of the comparator of each FE-I4 pixel was tuned individually using test pulse injection from the test pulse input of the CCPD_LF chip.

Fig. 5 shows the distribution of ToT values of one FE-I4 pixel. Test pulses were injected from the input of the CCPD_LF chip. When test pulses were injected from Pix1 in Fig. 1, the ToT values were 1 or 2. When test pulses were injected from Pix2, the ToT values were 3 or 4. All combinations were tested and are plotted in Fig. 5. The dispersion of ToT values is 1 or 2 and there is no overlap of ToT values. Therefore, the position of hits is possible to decode in the FE-I4 chip using the ToT values. Some pixels had overlapping ToT values because the pulse amplitudes of the output stages of the CCPD_LF are globally defined and only two parameters in the FE-I4 chip: the feedback current of

CSA and the threshold of the comparator can be tuned pixel-by-pixel to obtain the desired ToT values from the three output stages of the CCPD_LF chip. Additional tuning parameters would be required to completely tune the whole chip. However, the result has proved the feasibility of the concept of sub-pixel en/decoding.

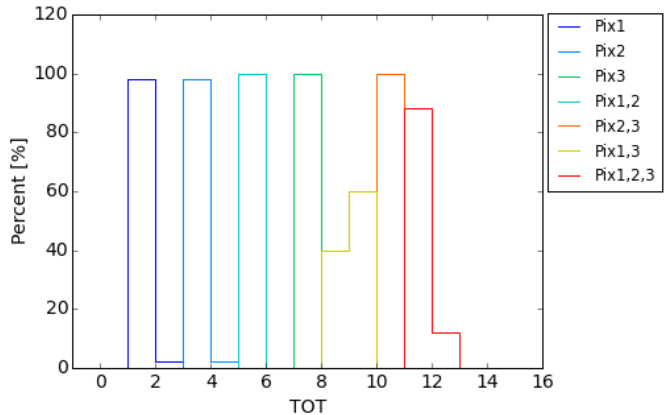

Fig. 5. Distribution of ToT values of the test pulses. Pix1, pix2, pix3 are corresponds to those in Fig. 1, respectively.

## VI. Conclusion

A depleted CMOS active pixel sensor prototype on a high resistive wafer was developed and thinned to 300 μm and 100 μm. The full depletion voltage of the 100-μm-thick chip is 20 V. The CCPD_LF chip operates sufficiently after irradiation with TID of 50Mrad and fluence of $1\times10^{15}$ $n_{eq}$/cm$^2$. However, degradation of the readout gain is as large as 30%. Fast readout was realized with an FE-I4 chip. One pixel of the FE-I4 chip was bump bonded to three pixels of the CCPD_LF chip. The signal position of the CCPD_LF chip is distinguishable in the FE-I4 chip using its ToT values.

With the success of this prototype, development of the next demonstrator chip is under way [9].

## References

[1] I. Peric, "A novel monolithic pixelated particle detector implemented in high-voltage CMOS technology", Nucl. Instr. Meth. A, vol 582, pp 876-885, 2007
[2] I. Peric, R. Eber, F. Ehrler, H. Augustin, N. Berger, S. Dittmeier, C. Graf, L. Huth, A.-K. Perrevoort, R. Phillipp, J. Repenning, D. vom Bruch, D. Wiedner, T. Hirono, M. Benoit, J. Bilbao, B Ristic and D. Muenstermann, "Overview of HVCMOS pixel sensors", JINST, 10, C05021, 2015.
[3] T. Hirono, M. Barbero, P. Breugnon, S. Godiot, L. Gonella, Tomasz Hemperek, F. Hügging, H. Krüger, J. Liu, P. Pangaud, I. Peric, D.-L. Pohl, A. Rozanov, P. Rymaszewski, A. Wang and N. Wermes," CMOS pixel sensors on high resistive substrate for high-rate, high-radiation environments", Nuclear Instruments and Methods in Physics Research Section A, vol. 831, pp 94-98, Sep. 2016.
[4] P. Rymaszewski, M. Barbero, P. Breugnon, S. Godiot, L. Gonella, T. Hemperek, T. Hirono, Fabian Hügging, H. Krüger, J. Liu, P. Pangaud, I. Peric, A. Rozanov, A. Wang, and N. Wermes," Prototype Active Silicon Sensor in 150 nm HR-CMOS Technology for ATLAS Inner Detector Upgrade", JINST, 11 C02045, 2016.
[5] ATLAS Collaboration, "CERN Letter of Intent for the Phase-II Upgrade of the ATLAS Experiment, Technical Report", CERN-LHCC-2012-022, LHCC-I-023, Dec. 2012.
[6] G. Lindström, M. Moll, E. Fretwurst, "Radiation hardness of silicon detectors – a challenge from high-energy physics", Nuclear Instruments and Methods in Physics Research Section A, vol. 426, pp 1-15, Apr. 1999.
[7] I. Peric, "Active pixel sensors in high-voltage CMOS technologies for ATLAS", JINST, 7 C08002, 2012.
[8] E. Vilella, M. Benoit, R. Casanova,G. Casse, D. Ferrere, G. Iacobucci, I. Peric and J. Vossebeld, "Prototyping of an HV-CMOS demonstrator for the High Luminosity-LHC upgrade", JINST, vol. 11 (01), C01012, 2016
[9] T. Wang, P. Rymaszewski, M. Barbero, Y. Degerli, S. Godiot, F. Guilloux, T. Hemperek, T. Hirono, H. Krüger, J. Liu, F. Orsini, P. Pangaud, A. Rozanov and N. Wermes "Development of a Depleted Monolithic CMOS Sensor in a 150 nm CMOS Technology for the ATLAS Inner Tracker Upgrade", JINST, submitted, Nov. 2016.
[10] M. Karagounis, Development of the ATLAS FE-I4 pixel readout IC for b-layer Upgrade and Super-LHC, in: Topical Workshop on Electronics for Particle Physics, Naxos, Greece, 15–19 CERN-2008-008, 2008, pp 7075. Sep. 2008,
[11] B.L. Henke, E.M. Gullikson, and J.C. Davis, "X-ray interactions: photoabsorption, scattering, transmission, and reflection at E=50-30000 eV, Z=1-92", Atomic Data and Nuclear Data Tables vol. 54, no.2, pp. 181-342, July 1993.
[12] M. Guthoff, O. Brovchenko, W. de Boer, A. Dierlamm, T. Müller, A. Ritter, et al., "Geant4 simulation of a filtered X-ray source for radiation damage studies", Nuclear Instruments and Methods in Physics Research Section A, vol. 675, pp.118–122, 2012.
[13] L. Snoj, G. Žerovnik and A. Trkov, "Computational analysis of irradiation facilities at the JSI TRIGA reactor", Appl. Radiat. Isot., vol. 70, pp. 483, 2012.
[14] https://github.com/SiLab-Bonn/pyBAR.